# MEDICAL THESES AND DERIVATIVE ARTICLES: DISSEMINATION OF CONTENTS AND PUBLICATION PATTERNS

**Mercedes Echeverria, David Stuart, Tobias Blanke**

## ABSTRACT

Doctoral theses are an important source of publication in universities, although little research has been carried out on the publications resulting from theses, on so-called derivative articles. This study investigates how derivative articles can be identified through a text analysis based on the full-text of a set of medical theses and the full-text of articles, with which they shared authorship. The text similarity analysis methodology applied consisted in exploiting the full-text articles according to organization of scientific discourse (IMRaD) using the TurnItIn plagiarism tool. The study found that the text similarity rate in the Discussion section can be used to discriminate derivative articles from non-derivative articles. Additional findings were: the first position of the thesis's author dominated in 85% of derivative articles, the participation of supervisors as coauthors occurred in 100% of derivative articles, the authorship credit retained by the thesis's author was 42% in derivative articles, the number of coauthors by article was 5 in derivative articles versus 6.4 coauthors, as average, in non-derivative articles and the time differential regarding the year of thesis completion showed that 87.5% of derivative articles were published before or in the same year of thesis completion.

## INTRODUCTION

Doctoral theses constitute an essential component of postgraduate education and an important source of scientific publications in universities. As a matter of definition, a thesis is an original research work "which presents the author's research and findings and submitted by him in support of this candidature for a degree or professional qualification" (ISO, 1986, p.679), an examination document necessary to obtain the doctoral degree (Paillassard et al., 2007), and an important piece for the transfer of scientific knowledge, economic and cultural development within the framework of the European Higher Education Area (Berlin Communiqué, 2003; Bergen Communiqué, 2005).

This study concentrates on doctoral theses in medicine, which represent an important resource of research. Publishing their results in peer-reviewed journals is an indicator of the scientific value of the research and makes the research known to scientific community for their scrutiny and validation.

A study of the Observatoire des Sciences et Techniques (2002) estimated that the research activity of PhD students represented 10-20% of indexed academic research within Scientific, Technical and Medical (STM) publications. A more recent study by Larivière (2012) determined that contributions from PhDs accounted for about a third of publication output in natural and medical sciences, although any relationship with their PhD thesis was not reported.

The research on doctoral theses, although recurrent in the literature, has traditionally been limited by three factors. Firstly, as Swales (1990) points out, the analysis of theses has been largely avoided by the size of the text. Secondly, the perceived lower quality or poor design of many dissertations (Diez et al., 2000; Brown, 2010; Bevan, 2005). Finally, it has traditionally been difficult to access the full-text of theses in university libraries (Macduff, 2009).

In the present work, a collection of medical theses was studied to determine how the content of doctoral theses is reused in the subsequent publication of scientific articles. Such a publication is often considered the logical step in the research process, as it both maximizes the impact of a piece of research and provides a means for a researcher to demonstrate its impact.

The theses examined corresponded to the traditional PhD form of a doctorate. In the UK, there are two distinct routes to a PhD, one based on research in a supervised programme and the other based on publications made independently of such a programme (UKCGE, 1998). The main distinction between these routes rests on their composition and size: the traditional thesis consists of a cohesive presentation in monograph, usually of about 80,000 words (Hoddell et al., 2002), while the PhD by publication consists entirely or predominantly of refereed and published articles in journals or books, which are already in the public domain (UKCGE, 1998), together with a usually 5,000-10,000 word critical appraisal to set the work in context (Davies et al., 2009)

This article aims to identify by means of textual analysis research the articles derived from medical theses, so-called *derivative works*. Although the meaning of such a term may be thought to be broadly understood, the parameters under which a work become 'derivative' are often not well-defined (Caan and Cole, 2012; Dhaliwal et al. 2010). Output and derivative works are semantically related terms, used sometimes as synonyms. However, the difference between both terms lies with the concept of authorship. An 'output' can be understood as any knowledge or result transferred from a work, whereas a 'derivative' is a new work based on a pre-existing original work produced only by the copyright owners (UK Copyright Service 2012). For our research, we define derivative works based on three factors that need to come together. A scientific article will be a derivative of a dissertation thesis if it is published by the same author, shares text and is similar in the content and produced during a doctorate or immediately after the completion of the thesis.

## Background

A number of quantitative studies have investigated publications generated from theses in medical journals. Their objective has been both to determine the number of publications from theses and to assess the quality of theses according to the impact factor of journals. The approach of most studies has identified the publications produced from theses in multidisciplinary databases (e.g. Web of Knowledge, Scopus), or medical databases (e.g. Medline, Lilacs, SciELO) or academic search engines (e.g. Google Scholar). Researchers have attempted to measure the prevalence of publications by means of text similarity based on short fragments, such as keywords of the title and abstract, combined with the name of the thesis's author, and occasionally, with the name of the thesis´s supervisors (Benotmane et al. 2012; Dhaliwal et al. 2010)

Salmi et al. (2001) performed a study on 300 medical theses submitted between 1993 and 1998 in French universities, and discovered that 17% of theses resulted in indexed publications. Benotmane et al. (2012)

determined that from 2150 TEM (Thèses d'Exercice en Médicine) defended in the Medical School of Lille 2, between 2001-2007, 243 TEM (11.3%) were object of a scientific publication, whilst also stating that "the real number of publications emanating from theses is not known and thus the comparisons to other schools of medicine of France or other countries were impossible" (Benotmane et al. 2012, p. e402). The same difficulty in determining the number of publications from doctoral degree programmes in the UK was raised by Caan and Cole (2012), when analysing the clinical research publications from doctoral research of 39 universities between 2000-2010. Results showed that 47.6% of theses produced no discernible publications.

In Germany, Cursiefen and Altunbas (1998) evaluated the contribution of medical doctors at the University of Würzburg in 1993-1995. They found that for 66% of medical students their doctoral research resulted in a Medline indexed publication. Whereas Ziemann and Oestmann (2012), who performed a retrospective study of the number of articles published by doctoral students at the Charité Universitätsmedizin, Berlin, in the decade 1998-2008, found that the number of doctorates who had published rose from 33% (1998) to 52% (2008). However, beyond this quantitative data Ziemann and Oetsmann raised the question whether parallel scientific projects for the doctoral period were documented as specific publications relating to the subject of a particular thesis. In this regard, a study on Spanish theses in anesthesiology, carried out by Figueredo et al. (2002), revealed that 204 doctoral degrees produced a total of 1679 articles, of which only 103 articles (6.1%) were related to a thesis.

A more nuanced approach was proposed by Dhaliwal et al. (2010) based on criteria of authorship and research outcomes between theses and publications in order to identity papers derived from a dissertation. Their analysis included 160 doctoral theses of a medical college in India. They found that the rate of papers in indexed journals was 30%. A different situation was presented by Frkovic et al. (2002) regarding scientific papers based on theses in two Croatian medical schools suggesting that the significant differences between publication rates of universities (Rijeka 11.1%, Zagreb 40.9%) were due to institutional emphasis directed towards publications.

Despite this wide range of studies, their results are not easily comparable. As Larivière indicates most of studies have been done small-scale case studies and "no study has yet attempted to measure the output of theses at a macro level." (Larivière 2012, p.465). This is not only because of different programmes of postgraduate systems, different lingual contexts where papers are not always indexed in databases, and different publication policies among institutions, but because it remains difficult to determine the published papers associated with PhD theses. Whilst the increase in open access policies means that the full-text of theses is now more widely available and analysis may be done with the full-text on a larger scale, there are necessarily methodological issues to be addressed; the most notable one being how derivative articles are identified.

## Research questions

This article pursues a systematic approach to determine whether derivative articles from theses can be identified through a text similarity analysis. Based on a full-text analysis of a set of biomedical theses compared with the full-text of articles, with whom they share authorship, this paper poses the following questions:
- What indexes of text similarity predict that an article is a derivative of a thesis?
- What publication patterns and authorship characterize the derivative articles?

# METHODOLOGY

This study examined 51 medical theses deposited in ERA (Edinburgh Research Archives) Repository of the Edinburgh University and 199 research articles published by the same theses authors. The ERA repository was selected as a representative example of a collection of biomedical theses, because it offers a sufficient homogeneous collection of theses, provides complete metadata records and a high productivity of authors as well as bibliometric tools. All of this means, that we can rely on a sufficient amount of information for our research. Furthermore, ERA has been a pioneering institutional repository (Jones & Andrew 2005; Jones et al. 2006) and can be considered as an example other repositories try to base their own efforts on. We can imagine that once more repositories are to the same standard that our study can be applied to a wide range of institutional repositories.

All theses investigated were completed between 2007 and 2011. In order to avoid possible biases in the dates of publications the time frame of articles was normalized. Only articles are considered that are not older than 2 years than the date of the thesis completion.

Research articles were identified by searching for the authors' names in two bibliographic databases: Scopus and Web of Knowledge. These databases were chosen due to their coverage of biomedical literature. Similar literature was found in both databases with no significant differences in number of papers per author. To address multiple authors by the same name we introduced an author disambiguation method using the available metadata of the author's affiliation. The analysis was focused on peer-reviewed articles because these contain a unified discursive structure IMRaD (Introduction, Methodology, Results and Discussion), which is required to measure the data distribution. Other types of documents (i.e. abstracts, clinical cases, editorials, meta-analysis and reviews) not based on this structure were excluded. In spite of examples of publications nine years before and two years after the thesis completion the majority of articles published fall within a period from 2.7 years mean (SD 2.0) before and 1.4 years mean (SD 0.5) after the thesis completion.

## Text similarity comparison

Text similarity was detected using the anti-plagiarism tool TurnItIn, a commercial software intended for verifying the originality of scholarly content. This program was chosen because it can process large quantities of text; the average medical thesis in this study contains 72,433 words and an article 7,428 words. In addition, TurnItIn was successfully used before to provide a quantitative approach to performing a textual analysis of published works (Sun, 2013).

For the purpose of our study, we needed a program that would fulfil the following requirements: it would analyse large textual corpus, it would operate intra-corpal, so that the source and copy can be both in a corpus, it would give a numerical measure of the similarity of two documents as a 'similarity score', and it would present all matches in detail, indicating their location in the text. According to this description the only programs that could offer these functionalities were Ithenticare and Turnitin, both of which share the same technology, database content and belong to the same company, iParadigm. We finally decided on TurnItIn after comparing it with other programmes such as: eBLAST (with optimum input 50-500 words), EVE2 (papers), CopyCatch (essays) and JPlag (plagiarism).

The full-text of the 51 theses and 199 articles was uploaded to TurnItIn. This software highlights the position where in the text similarity matches occur and generates a similarity index between texts. Furthermore, as Dreher (2007) confirmed in studies on plagiarism, the tool offers a very reasonable performance when the scope of analysis is constrained to a controlled corpus.

TurnItIn detects similarity at the text-string level. String-matching procedures aim to identify identical text strings that are treated as indicators for potential association between documents. The algorithm underlying TurnItIn is not publicly known. However, experiments with TurnItIn carried out by Introna (2007) detected that the algorithm operates on the basis of creating a digital 'fingerprint' of a document, which is used to compare documents with each other. The detection of similarities by TurnItIn is dependent on a range of characteristics: long strings of consecutive words from the original text retained in the copied version, a small amount of variation of the order of words and small changes of consecutive strings within a fragment. Known limitations of TurnItIn are linked to the recognition of lexical patterns. It therefore does not detect paraphrases, translations or the adoption of ideas.

**Cluster methodology**

The primary objective of the research consisted of gathering statistical data of textual similarity between theses and articles. TurnItIn generates its similarity index as a percentage based on a summary of matching similar text found in the document submitted (articles) against the target document (the thesis of the same author of articles). In order to understand the value of this index we explored the degree of dependence between the TurnItIn similarity index and the total of 'matches' produced in each document submitted using Spearman's correlation analysis for non-parametric variables, because the data did not present a normal distribution pattern.

In this paper, the approach consisted of measuring the text similarity using the IMRaD standardised structure (Introduction, Methodology, Results and Discussion) to exploit the full-text of articles according to the organization of the scientific discourse in it. IMRaD was introduced for the health articles in the 1940s (Sollaci and Pereira, 2004) and it is the format accepted for most of biomedical articles (ICMJE, 2010). IMRaD provides a structure for the organization of research and argumentation of articles, which involves different communicative purposes and discursive functions for each rhetorical category.

The method consisted of clustering the text articles according to their structure in different sections (title, abstract, introduction, methodology, results and references) in order to facilitate the analysis as well as the comparison and interpretation of data belonging to similar discursive units. References were not included in the analysis because they are not part of the scientific originality of articles.

Subsequently, we wanted to know the distribution of text similarity along the different sections of articles with the aim of providing evidence about which sections presented the highest levels of text similarity when compared with the original theses. For that, we computed the data of matches of similarity for each section.

Given that the starting point was to cluster the articles according to their textual structure, a proximity matrix was used for measuring the distance between all pairs of clusters regarding the TurnItIn similarity index. This approach provides insights into which discursive sections are responsible for a similarity between theses and articles. The shortest distance with regard to similarity indicates which sections have the highest similarity and the furthest distance the lowest similarity.

A question more specific derived from matrix analysis would consist in determining the values of proximity and distance between the different sections of articles. This approach would provide insights into which discursive sections are closed and how they are related to each other.

To understand how to perform the different sections of the articles regarding the values of similarity a dendrogram was created from the proximity matrix. The dendrogram was constructed using the average linkage among groups, which uses as distance the average of clusters. This dendrogram displays how strongly the individual sections of articles are correlated, based on their degree of similarity.

The final objective of our research was to identify an optimal cut-off point within discursive sections with high levels of similarity that could determine whether an article is classified as a derivative. For this, we used the Receiver Operating Characteristic (ROC) curve for estimating the discrimination of derivative from non-derivative works. This curve has been used in medical decision-making, and recently in data mining research (Fawcett 2006). The ROC curve is a two-dimensional graph that visually depicts the full picture of trade-offs between the sensitivity (proportion of true positives) and 1-specificity (proportion of false positives) across a series of cut-off points in order to identify correctly the optimal threshold point that could predict the derivative works. The sensibility and specificity are two indicators that depend on the value of reference (gold standard). Within this study the 'gold standard' included the articles bound in the theses and those articles that the authors mentioned as publications produced from theses. Examples include: "The work discussed in Chapter 3 has been published in (…)", "List of publications arising from this thesis", "Publications from this thesis", "Publications resulting from the work presented in this thesis"; or with less explicit information "The following papers were published during the course of this thesis".

Subsequent questions arising from the research included authorship, the allocation of authorship credits and co-authorship, as well as how the supervisors of theses are represented in the derivative articles. While there is no consensus on quantifying authorship credit, we used the harmonic credit designed by Hagen (2010) to quantify the credit that can be attributed to authors. Hagen offers a model that gives credit according the position, which authors occupy. This seemed a reasonable formula for our domain of research, as it assigns credits according to the position of authors in the authorship list. The formula used was:

$$\text{ith author credit} = (1/i)/[1+ (1/2) + \ldots + (1/N)]$$

In our experiments, the statistical significance was set at 5%. All our tests were two-sided and the analyses were carried out using the SPSS software (version 20).

# RESULTS

## Full-text and cluster analysis

The first analysis step consisted of defining the groups. Seven different clusters were generated according to the sections of articles (Title, Abstract, Introduction, Methodology, Results, Discussion and References). An Excel spreadsheet was used to compute the percentage of the TurnItIn similarity index between thesis and articles, the distribution of matches along sections and the total number of matches per article (Table 1).

| AUTHORS-ARTICLES | SIMILARITY INDEX | TITLE | ABSTRACT | INTRODUCTION | METHODOLOGY | RESULTS | DISCUSSION | REFERENCES | MATCHES |
|---|---|---|---|---|---|---|---|---|---|
| Author1-Article1 | 3% | 0 | 0 | 0 | 1 | 4 | 0 | 9 | 14 |
| Author2-Article1 | 10% | 0 | 0 | 4 | 24 | 42 | 4 | 20 | 94 |
| Author2-Article2 | 0% | 0 | 0 | 0 | 0 | 0 | 0 | 0 | 0 |
| Author3-Article1 | 61% | 1 | 6 | 21 | 43 | 38 | 30 | 36 | 175 |
| Author4-Article1 | 6% | 0 | 0 | 0 | 0 | 0 | 0 | 15 | 15 |
| Author4-Article2 | 42% | 1 | 1 | 14 | 29 | 57 | 35 | 54 | 191 |
| Author4-Article3 | 7% | 1 | 2 | 0 | 0 | 0 | 0 | 15 | 18 |
| Author4-Article4 | 5% | 1 | 1 | 0 | 3 | 0 | 1 | 0 | 6 |
| Author5-Article1 | 2% | 0 | 0 | 0 | 8 | 0 | 0 | 10 | 18 |
| Author5-Article2 | 2% | 0 | 0 | 0 | 5 | 0 | 0 | 3 | 8 |

(Table 1: Example of the spreadsheet format. Text similarity and matches between theses and articles. Theses authors are numbered and their articles are also numbered)

TurnItIn generates a similarity index, which indicates the degree to which an article matches the thesis. In order to know the degree of dependence between both variables, Similarity and Matches, the Spearman correlation coefficient was calculated. The coefficient was 0.953, with a statistical significance $P<0.001$, which showed a strong correlation between both variables.

With the aim of studying the distance of the sections of articles regarding the TurnItIn similarity index as well as the distance and relationships between each pair of sections, a proximity matrix was constructed. The specifications were Euclidean distance squared and the initial values were transformed by the type, range 0- 1 (Table 2).

| Case | Matrix File Input | | | | | | |
|---|---|---|---|---|---|---|---|
| | SIMILARITY INDEX | TITLE | ABSTR. | INTRODUC | METHODOL. | RESULTS | DISCUSSION |
| SIMILARITY INDEX | .000 | 10.310 | 6.740 | 5.608 | 4.474 | 4.774 | 4.720 |
| TITLE | 10.310 | .000 | 7.326 | 7.910 | 8.085 | 8.095 | 7.377 |
| ABSTRACT | 6.740 | 7.326 | .000 | 4.443 | 5.428 | 5.318 | 3.674 |
| INTRODUCTION | 5.608 | 7.910 | 4.443 | .000 | 5.464 | 4.749 | 3.193 |
| METHODOLOGY | **4.474** | 8.085 | 5.428 | 5.464 | .000 | 4.289 | 3.684 |
| RESULTS | 4.774 | 8.095 | 5.318 | 4.749 | 4.289 | .000 | 2.896 |
| DISCUSSION | 4.720 | 7.377 | 3.674 | 3.193 | 3.684 | **2.896** | .000 |

(Table 2: Distance matrix of distances between clusters)

The matrix showed that the section closest to similarity index were Methodology (4.474). The low value of Methodology can be due to several factors: similarity in a methods section is quite common, as methods are often fairly standardised in disciplines (Sun et al., 2010), Methodology is a commonly re-used and shared section and changes in methodologies occur slowly in the scientific research. Regarding the distances between the different discursive sections of the articles, Discussion and Results showed the closest distance (2.896), which could indicate an internal linkage of the discourse organization.

To obtain a picture of the distances between the sections a dendrogram was created (Figure 1) that provides the rescaled distance how the clusters are combined.

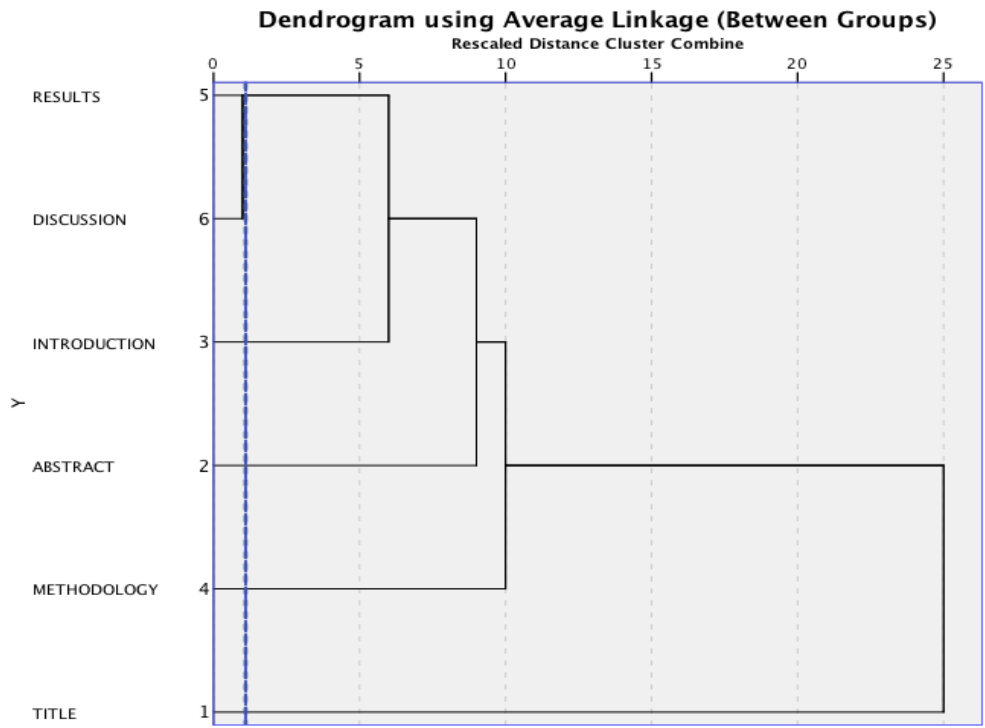

(Figure 1: Hierarchical cluster dendrogram of the distances between the sections of articles)

The dendrogram in Figure 1 is comprised of five clusters. The most significant clusters is the first, taking into account the criteria of distance. As can be observed, the first agglomeration (Results-Discussion) is a cluster clearly defined and homogeneous. Results and Discussion are the sections where the novelty and originality of article is present. From a rhetorical perspective, both sections are interdependent. Discussion describes the content of the Results section and generally emphasizes their accuracy and consistency.

The clusters (Introduction-Abstract-Methodology) represent small distances between them, the discrimination of small distances is not significant. They are statistically blurred groups, heavily correlated, where it is complex to identify categories. In fact, small variations in the data distances could yield very different conglomerates, whereas the cluster (Result-Discussion) would remain stable. In rhetorical terms, the concepts (Introduction-Abstract-Methodology) have much in common, they are sections that are mutually related, for which it is then difficult to assign a rating statistically.

**Statistical identification of derivative works**

In order to discriminate derivative articles from non-derivative articles a ROC curve analysis was applied. 37 articles were identified as a 'gold standard' measure. The ROC curve estimated the optimal threshold point to predict derivative works (Figure 2).

The area under the curve (AUC) corresponds to the probability of identifying derivative and non-derivative articles within two dichotomous variables. The larger is AUC the better is overall performance of the test.

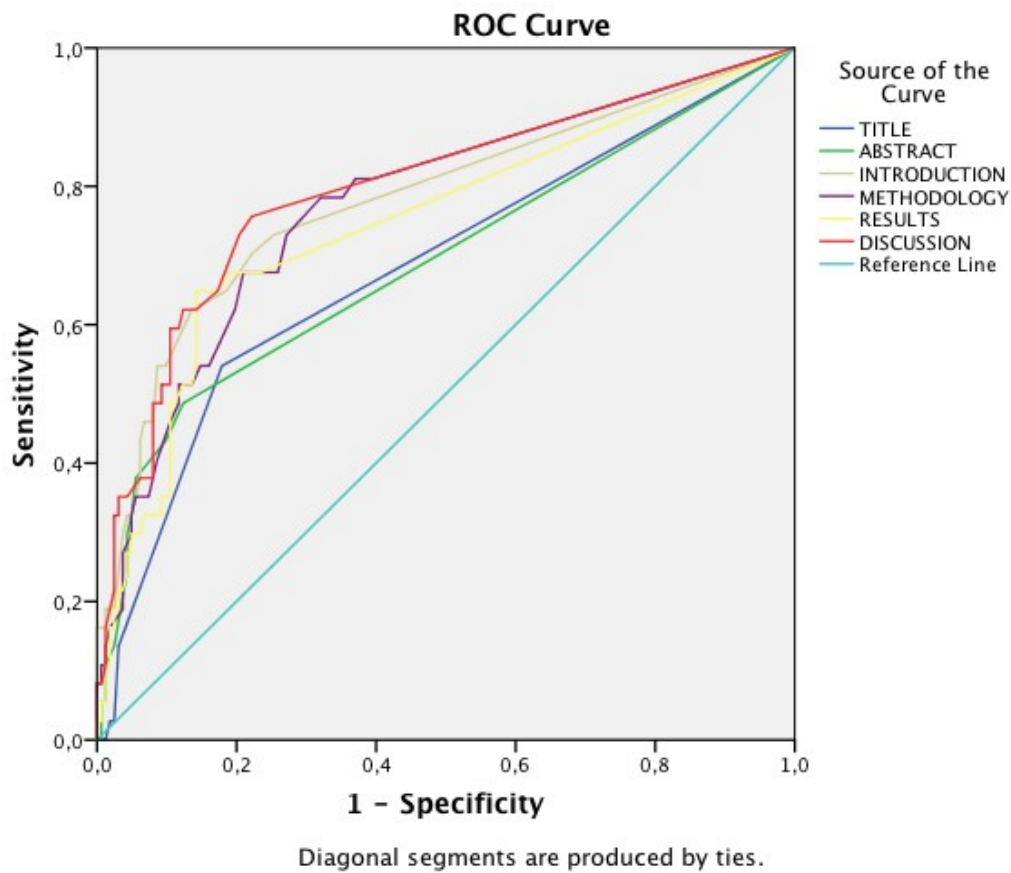

(Figure 2. ROC curve of derivative articles. Area under curve= **0.796, Discussion**)

The results showed that the greater AUC corresponded to the Discussion section with 0.796 ([95% CI], **0.70-0.89**), which suggests that Discussion is very predictable with respect to the identification of derivative articles.

The method used to calculate the optimal threshold point was the Youden Index (Kumar and Indrayan 2011), obtained by deducting 1 from the sum of test's sensitivity and specificity expressed not as percentage but as a part of a whole number.

The cut-point calculated in Discussion was 7.5 (sensitivity 0.60 and 1-specificity 0.098), being 0.098 false positives (FP). This cut-point was used to discriminate derivative articles from non-derivative articles. For the validation of our data, please compare (Appendix 1. Table 1).

## **Overall**

The number of articles identified as derivatives using the cut-point in the Discussion of 7.5 **were 40 out of 199, 20.1 %.**
Not all articles bound in theses turned out to be derivative articles, 15 of them were classified as non-derivatives. Furthermore, 18 articles not included as part of the original 'gold standard' were derivative articles.

The similarity index of derivative articles ($N$=40) regarding theses was **39.7%**, as mean,
17.2 SD) versus **5.3 %,** as mean (9.8 SD) in non-derivate articles ($N$=159).

Other indicators that support the effectiveness of our method in order to identify derivative peer-review articles from a thesis were:

1) *The thesis author as first author of articles*
    a) The thesis author as first author of derivative articles was accounted in 34 out of 40 articles, **85%.**
    b) The thesis author as first author of non-derivative articles was accounted in 51 out of 159 articles, **32%.**

2) *Allocation of authorship credit*
To calculate the authorship credit allocation the formula designed by Hagen (2010) was used: ith author credit = $(1/i)/[1+ (1/2) + \ldots + (1/N)]$. Our results were:

    a) The authorship credit mean of derivative articles was 0.42, which is equivalent to the position of author thesis in the article **1/4.**
    b) Whereas in non-derivative articles the authorship credit mean was 0.24, which is equivalent to the position of thesis author **2/4**.

3) *Co-authorship and thesis supervisors*

    a) For derivative articles in 40 (**100%**), supervisors collaborated within the range of our investigation:
1/1 supervisor = 13/40 articles
1/2 supervisors = 7/40 articles
2/2 supervisors = 18/40 articles
2/3 supervisors = 1/40 articles
3/3 supervisors = 1/40 articles

    b) For 106 out of 159 non-derivatives articles (**66.7 %),** supervisors also collaborated within the range:
1/1 supervisor = 36/106 articles
1/2 supervisors = 38/106 articles
2/2 supervisors = 23/106 articles
2/3 supervisors = 4/106 articles
3/3 supervisors = 5/106 articles

    c) In 53 of 159 non-derivative articles (**33.3%**), we found co-authors that were not supervisors:
0 supervisors = 53/159 articles.

The data showed that all derivative articles (*N*=40) had supervisors as coauthors with a mean of the TurnItIn similarity index of 39.7% (see Figure 3). Non-derivative articles, whose coauthors were supervisors, had a mean TurnItIn similarity index of 7.5% (see Figure 4), and non-derivative articles whose coauthors were not supervisors had a mean similarity index of 0.94% and therefore statistically non-relevant (see Figure 5). The data can be validated in (Appendix I. Tables 1, 2 & 3).

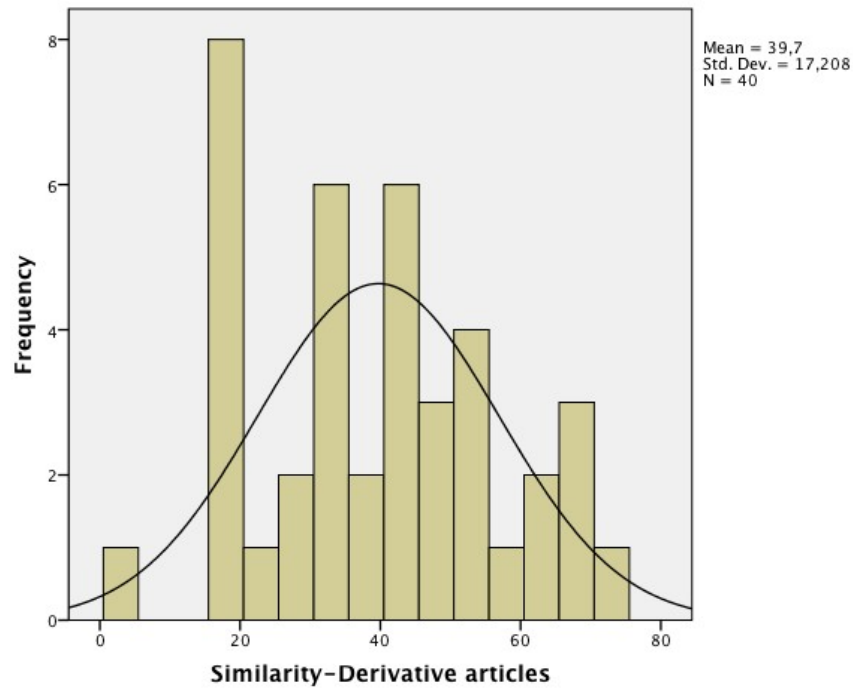

(Figure 3: Derivative articles with supervisors as co-authors. Similarity index 39.7% mean)

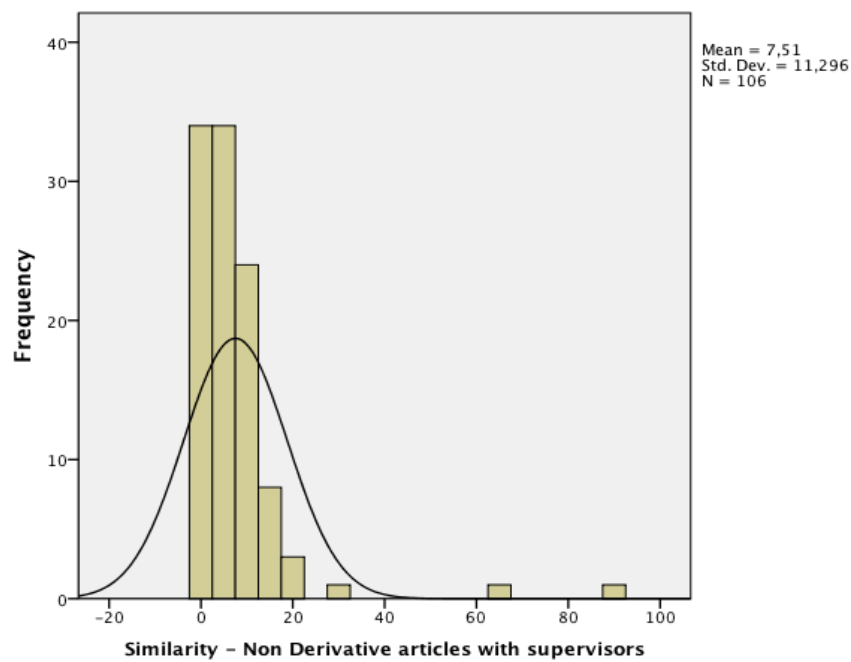

(Figure 4: Non derivative article whose co-authors were supervisors. Similarity index, 7.5% mean)

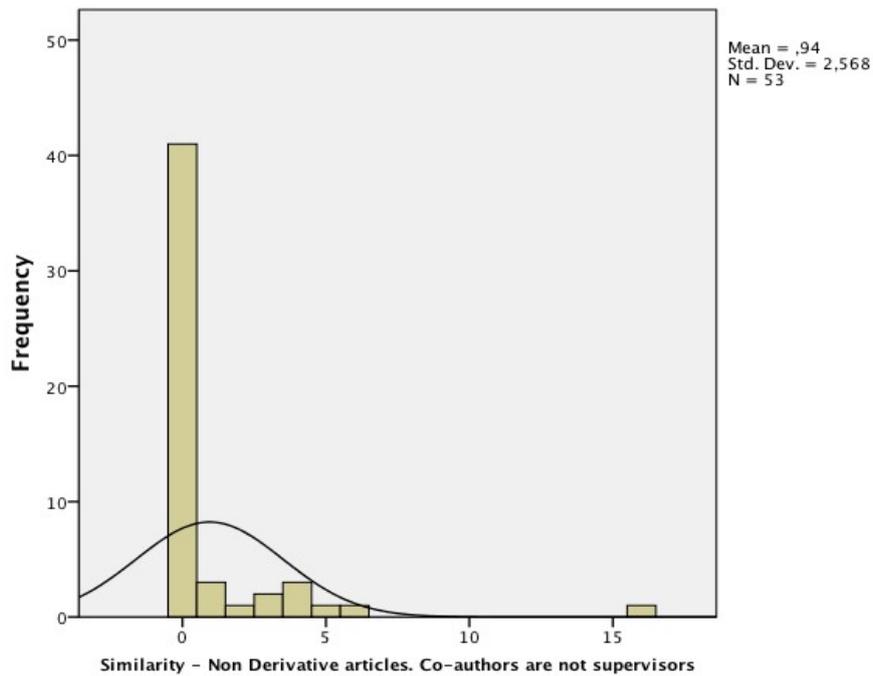

(Figure 5: Non derivative article whose co-authors were not supervisors.
Similarity index 0.94 % mean)

4) <u>Number of authors by article</u>:
  a) Derivative articles: 5 authors/article (SD 2.0).
  b) Non-derivative articles: 6.4 authors/article (SD 3.5)

5) <u>Time differential of articles regarding year of thesis completion</u>
  a) Derivative articles published before or in the same year of thesis completion: 35/40 (87.5%)
  b) Non-derivative articles published before or in the same year of thesis completion 93/159 (58.5%)

# DISCUSSION

Our method of using a full-text similarity index to compare theses and articles turns out to be useful for identifying derivative articles based on thesis, published by the same author of thesis.

The analysis demonstrated that a high textual similarity in the Discussion section determines a strong correlation between theses and derivative articles, which is the essential component for the association of both documents. This finding is consistent with Sun et al. (2010) who pointed out that "The probability of hight abstract similarity given similar results/Discussion sections is significantly higher than the probabilities of high abstract similarity given similar methods sections…because the novelty of research articles is typically demostrated more in its results/Discussion sections".

Studies of text similarity in abstracts and titles have provided interesting approaches to detect papers derived from theses (Salmi et al. 2001, Benotmane et al. 2012, Caan et al. 2012), although they have not been able to discern with clear evidence outputs emanating from theses. However, on the basis of full-text comparison of theses and articles of the same author, the Discussion section has demostrated to be the essential part to identify the derivative articles. From a cut-off (7.5) of text similarty obtained in Discussion, it was possible to distinguish the derivative ($N$=40/199) from non-derivative ($N$= 159/199) articles.

With regard to text similarity, the derivative articles had a mean similarity of 39.7% (17.2 SD) with theses while non-derivative articles had 5.3% as mean, (9.8 SD) using the TurnItIn software for quantitatifying the textual comparison. Complementary evidence on derivative articles can be based on a high similarity index reached by Results. The data confirmed the proximity and interdependence of the Discussion and Results sections, where the most scientific evidence of articles can be found.

In terms of authorship the order of authors showed the prevalence of a thesis's author as also the first author in 85% (34/40) of the derivative articles versus 32% (51/159) in non-derivative ones. These results reflect the general premise by which the name of the principal investigator is almost always mentioned first (Subramanyam 1983). However, other studies conducted on outputs of theses showed inconsistent results. While Arriola-Quiroz et al. (2010) reported that medical students were first authors of the articles in 83.5% of cases, Dhaliwal et al. (2010) identified postgraduate student as first authors in 54% of papers based on theses. Diez et al. (2000) finally argued that regarding medical theses in a German medical faculty, students were first authors for 8% of Medline indexed papers. Thus, these investigations confirm that the order by which authors are listed on a paper is a complex topic and one of the least standardized ones (Jones and McLellan 2000). As it is indicated by the ICMJE (2010) "the decision of the order of authorship is to the coauthors".

Regarding the number of authors of derivative articles (5 mean), there are discernible differences to the average number of authors per article in biomedical journals 6.9 (Weeks et al. 2004) or 6.69 (Costas et al., 2011). These differences could be linked to two factors: *the authorship credit*, determined by the contribution of the thesis's author in a proportion of 42%, as average of derivative articles, and the type of collaboration of PhD students in the academic research.

The participation of PhD students in research teams has already been noted by Larivière (2012) as a determining factor for science students. The collaboration of PhD students in research groups gives them the opportunity to publish with advisors, supervisors, mentors and other members of team. As Bordons et al. (2000) pointed out, there are *intra-teams and extra-teams collaboration* in scientific networks. According to data of this study it seems that the integration of PhD students occurs at intra team level, with a high rate of collaboration (73.3%) between supervisors and of PhDs students in the publications. This evidence is consistent with the findings of Pole, mentioned by Larivière (2012) " [even] if we do not have any information of the link between the student and other authors, it could be expected, that those co-authors were supervisors and mentors". This sociological aspect of research governed by personal interactions will be highly relevant with regard to the number of authors of derivative articles.

More specifically, a finding of our research was the participation of supervisors as the essential component for derivative articles. The analysis demonstrated that all 40 derivative articles had supervisors as co-authors. In fact, it could be shown that a difference between derivative and non-derivative articles is the participation of supervisors as co-authors. In particular, the articles written by PhD students without collaboration of supervisors presented an irrelevant TurnItIn similarity index (0.94%) with theses.

Another general feature that emerges from our analysis was the correlation between the date of publication of derivative articles and the date of the theses' completion. The data evidenced that 35/40 (87.5%) of derivative articles were published before or during the year of thesis submission versus 93/159 (58.5%) of non-derivative articles. This correlation between derivative articles and date of publication appears to be linked to several factors: reliability and validity of peer-review, which ensures rigorous evaluation and increases the quality of thesis, the entrance of PhDs to academic careers and institutional policies for doctoral degrees.

Regarding the timing of publications relating to theses, we found similar results in the study presented by Caan et al. (2012) for 39 universities in the UK. They demonstrated with a sample of 82 theses, that 43 theses (52.4%) produced publications related to doctoral research, whose publication year was "preceded the award in 29 cases (mean interval 2.7 years earlier), six cases published in the same year and eight published after the award (mean interval 3.0 years later)". That means, on average, that 81.3% of authors published related papers before or during the year of the thesis completion. Other authors reported publications of medical students using different criteria, such as: Salmi et al. (2001) on French medical theses, with any article registered before the thesis submission, 14 papers (27.5%) in the year of presentation, with a cumulative proportion of 49.0% of papers having been published after 2 years. Dhaliway et al. (2010) on medical theses in an Indian university did not include the possibility of articles derived before or during the year of the thesis completion, and Larivière (2012) in a study about the contribution of Canadian PhD students in peer reviewed students revealed that 63% of doctoral students in health contributed to at least one paper during their doctorate, although any relationship with their PhD thesis was reported.

Finally, the results obtained in this work were consistent with other empirical studies on derivative articles of medical theses. The percentage of derivative articles published by students during the doctoral studies and postdoctoral period in our study was 20.1%. This data is comparable to a survey by Pitche et al. (2007) among faculties, where "theses represented only 21.6% of these publications" [articles in indexed journals] and Baufreton et al. (2012) that conducted searches in databases with surveys of validation among theses supervisors and found that 165 theses from the School of Medicine of Angers, France, produced 28% of publications, out of 97 (16%) indexed in PubMed.

**CONCLUSION**

This paper examined a range of issues related to derivative articles from medical theses, a subject that is, to our knowledge, currently under-investigated in the literature. The paper studied derivative articles using textual analysis. We found that the Discussion section is highly predictive for identifying derivative articles based on text analysis. Furthermore, the paper analysed the publication patterns of derivative articles, whose most relevant characteristics in this respect are: a high prevalence of a thesis's author as the first author of derivative articles, the participation of supervisors as co-authors, a discernible difference in the number of authors and a high rate of publications during the doctoral studies.

Further research would be necessary to determine whether these findings could be extrapolated, not only by the size of sample examined, but also by disciplinary differences, as different scientific fields have developed diverse profiles in terms of research production, modalities of publication and working practices by scientists and scholars of different countries.

This analysis could have potential implications to assess the PhDs publications and provide insights into the scientific scope of theses as an important resource of information on the academic publications.

The study was focused on peer-reviewed articles due to their unified discursive structure required to measure the distribution of data, although it involves a limitation to exclude other potential types publications done by PhD students to disseminate their theses. Furthermore, it would be desirable for future studies to redefine the concept of a gold standard. Finally, interesting information would have been to get a feedback of the thesis authors. This complementary information might have provided to our study with a realistic view of PhD students research on their careers.

**Acknowledgments.** The authors acknowledge the assistance and suggestions of José Javier Sánchez-Hernández, Department of Preventive Medicine and Public Health, School of Medicine, Universidad Autonoma de Madrid, Spain.

**APPENDIX 1: VALIDATION OF DATA**

| AUTHORS-ARTICLES | SIMILARITY INDEX | TITLE | ABSTRACT | INTRODUCTION | METHODOLOGY | RESULTS | DISCUSSION | REFERENCES | MATCHES | AUTHOR POSITION | SUPERVISORS |
|---|---|---|---|---|---|---|---|---|---|---|---|
| Author3-Article1 | 61% | 1 | 6 | 21 | 43 | 38 | 30 | 36 | 175 | 1/7 | 2/2 |
| Author4-Article2 | 42% | 1 | 1 | 14 | 29 | 57 | 35 | 54 | 191 | 1/3 | 2/2 |
| Author5-Article4 | 21% | 0 | 2 | 6 | 43 | 40 | 36 | 54 | 181 | 1/8 | 2/2 |
| Author7-Article1 | 35% | 0 | 0 | 42 | 0 | 0 | 9 | 5 | 56 | 1/2 | 1/2 |
| Author7-Article2 | 54% | 0 | 0 | 0 | 31 | 26 | 37 | 46 | 140 | 1/6 | 2/2 |
| Author7-Article9 | 63% | 1 | 8 | 10 | 31 | 19 | 35 | 56 | 160 | 1/8 | 2/2 |
| Author8-Article1 | 20% | 0 | 0 | 2 | 20 | 29 | 43 | 47 | 141 | 1/7 | 1/2 |
| Author9-Article7 | 17% | 0 | 1 | 0 | 12 | 10 | 8 | 19 | 50 | 1/4 | 1/1 |
| Author10-Article2 | 34% | 3 | 1 | 1 | 49 | 71 | 11 | 132 | 268 | 1/11 | 3/3 |
| Author10-Article4 | 44% | 1 | 2 | 3 | 56 | 14 | 15 | 18 | 109 | 1/8 | 2/3 |
| Author11-Article1 | 33% | 0 | 0 | 16 | 20 | 38 | 28 | 24 | 126 | 1/4 | 1/1 |
| Author14-Article2 | 41% | 0 | 0 | 7 | 13 | 29 | 61 | 46 | 156 | 1/5 | 2/2 |
| Author14-Article3 | 43% | 0 | 7 | 3 | 32 | 33 | 20 | 16 | 111 | 1/4 | 2/2 |
| Author15-Article6 | 18% | 1 | 6 | 4 | 4 | 0 | 7 | 13 | 35 | 1/4 | 2/2 |
| Author15-Article8 | 32% | 4 | 3 | 9 | 32 | 0 | 27 | 23 | 98 | 1/3 | 2/2 |
| Author18-Article2 | 19% | 0 | 0 | 5 | 0 | 21 | 8 | 20 | 54 | 1/7 | 1/1 |
| Author19-Article2 | 47% | 0 | 6 | 17 | 28 | 41 | 41 | 69 | 202 | 1/2 | 1/1 |
| Author21-Article1 | 30% | 1 | 0 | 15 | 18 | 45 | 18 | 40 | 137 | 1/2 | 1/2 |
| Author22-Article11 | 35% | 0 | 0 | 11 | 11 | 37 | 21 | 21 | 101 | 1/3 | 1/1 |
| Author22-Article14 | 34% | 0 | 0 | 9 | 18 | 45 | 24 | 37 | 133 | 1/6 | 1/1 |
| Author22-Article15 | 69% | 1 | 10 | 19 | 24 | 39 | 20 | 34 | 147 | 1/8 | 1/1 |
| Author24-Article4 | 53% | 1 | 2 | 10 | 22 | 95 | 69 | 44 | 243 | 1/6 | 2/2 |
| Author24-Article6 | 50% | 1 | 0 | 21 | 29 | 82 | 18 | 53 | 204 | 1/5 | 1/2 |
| Author24-Article7 | 66% | 1 | 7 | 0 | 20 | 138 | 21 | 40 | 227 | 1/6 | 1/2 |
| Author27-Article1 | 54% | 1 | 12 | 34 | 0 | 10 | 43 | 13 | 113 | 2/2 | 1/1 |
| Author28-Article3 | 40% | 0 | 0 | 7 | 9 | 21 | 33 | 31 | 101 | 1/3 | 2/2 |
| Author28-Article4 | 73% | 2 | 15 | 31 | 60 | 92 | 98 | 77 | 375 | 1/4 | 2/2 |
| Author29-Article3 | 42% | 1 | 6 | 23 | 30 | 19 | 29 | 39 | 147 | 1/4 | 2/2 |
| Author30-Article3 | 67% | 1 | 9 | 14 | 86 | 60 | 30 | 48 | 248 | 2/3 | 1/1 |
| Author31-Article2 | 45% | 0 | 1 | 11 | 50 | 11 | 16 | 33 | 122 | 1/7 | 2/2 |
| Author34-Article2 | 18% | 0 | 5 | 14 | 17 | 44 | 13 | 11 | 104 | 2/4 | 1/2 |
| Author36-Article2 | 53% | 0 | 4 | 0 | 58 | 47 | 36 | 43 | 188 | 1/6 | 2/2 |
| Author37-Article5 | 59% | 1 | 6 | 15 | 26 | 70 | 63 | 68 | 249 | 1/6 | 2/2 |
| Author38-Article2 | 20% | 0 | 3 | 8 | 18 | 13 | 11 | 30 | 83 | 1/5 | 1/2 |
| Author44-Article1 | 30% | 0 | 2 | 13 | 22 | 93 | 21 | 54 | 205 | 1/4 | 2/2 |
| Author44-Article3 | 36% | 1 | 2 | 16 | 7 | 113 | 25 | 40 | 204 | 2/5 | 2/2 |
| Author46-Article3 | 50% | 0 | 0 | 11 | 46 | 0 | 16 | 30 | 103 | 1/2 | 1/1 |
| Author46-Article5 | 3% | 0 | 0 | 0 | 4 | 2 | 15 | 7 | 28 | 2/5 | 1/1 |
| Author51-Article1 | 17% | 0 | 0 | 2 | 2 | 12 | 19 | 18 | 53 | 2/5 | 1/1 |
| Author51-Article3 | 20% | 0 | 7 | 18 | 12 | 49 | 21 | 20 | 127 | 1/6 | 1/1 |

(Table 1: Derivative articles. Cut-off point in Discussion => 7.5)

| AUTHORS-ARTICLES | SIMILARITY INDEX | TITLE | ABSTRACT | INTRODUCTION | METHODOLOGY | RESULTS | DISCUSSION | REFERENCES | MATCHES | AUTHOR POSITION | SUPERVISORS |
|---|---|---|---|---|---|---|---|---|---|---|---|
| Author1-Article1 | 3% | 0 | 0 | 0 | 1 | 4 | 0 | 9 | 14 | 2/6 | 3/3 |
| Author2-Article1 | 10% | 0 | 0 | 4 | 24 | 42 | 4 | 20 | 94 | 1/8 | 1/1 |
| Author4-Article1 | 6% | 0 | 0 | 0 | 0 | 0 | 0 | 15 | 15 | 3/8 | 2/2 |
| Author4-Article3 | 7% | 1 | 2 | 0 | 0 | 0 | 0 | 15 | 18 | 3/4 | 1/2 |
| Author4-Article4 | 5% | 1 | 1 | 0 | 3 | 0 | 1 | 0 | 6 | 1/4 | 1/2 |
| Author5-Article1 | 2% | 0 | 0 | 0 | 8 | 0 | 0 | 10 | 18 | 6/8 | 1/2 |
| Author5-Article2 | 2% | 0 | 0 | 0 | 5 | 0 | 0 | 3 | 8 | 3/8 | 1/2 |
| Author5-Article3 | 2% | 0 | 0 | 0 | 7 | 0 | 0 | 10 | 17 | 5/8 | 1/2 |
| Author7-Article3 | 1% | 0 | 0 | 0 | 0 | 0 | 0 | 2 | 2 | 1/2 | 1/2 |
| Author7-Article5 | 1% | 0 | 0 | 0 | 5 | 0 | 0 | 0 | 5 | 3/5 | 1/2 |
| Author7-Article6 | 11% | 0 | 0 | 3 | 13 | 0 | 0 | 39 | 55 | 4/7 | 1/2 |
| Author7-Article7 | 0% | 0 | 0 | 0 | 0 | 0 | 0 | 0 | 0 | 2/6 | 1/2 |
| Author9-Article1 | 15% | 1 | 0 | 0 | 15 | 0 | 0 | 35 | 51 | 1/3 | 1/1 |
| Author9-Article2 | 12% | 0 | 2 | 0 | 12 | 36 | 0 | 35 | 85 | 1/6 | 1/1 |
| Author9-Article3 | 7% | 0 | 0 | 2 | 8 | 2 | 0 | 15 | 27 | 1/2 | 1/1 |
| Author9-Article4 | 6% | 0 | 0 | 0 | 11 | 0 | 5 | 26 | 42 | 1/9 | 1/1 |
| Author9-Article8 | 14% | 0 | 0 | 4 | 12 | 8 | 5 | 10 | 39 | 1/2 | 1/1 |
| Author10-Article1 | 3% | 0 | 0 | 0 | 7 | 1 | 0 | 9 | 17 | 6/12 | 2/3 |
| Author10-Article3 | 3% | 0 | 0 | 1 | 2 | 0 | 0 | 13 | 16 | 2/4 | 2/3 |
| Author10-Article5 | 5% | 1 | 4 | 1 | 5 | 0 | 0 | 16 | 27 | 8/8 | 2/3 |
| Author10-Article6 | 5% | 1 | 0 | 0 | 22 | 0 | 2 | 11 | 36 | 1/6 | 2/3 |
| Author11-Article2 | 8% | 1 | 0 | 0 | 15 | 5 | 0 | 29 | 50 | 1/6 | 1/1 |
| Author11-Article3 | 8% | 0 | 0 | 4 | 12 | 5 | 2 | 17 | 40 | 2/5 | 1/1 |
| Author12-Article2 | 12% | 2 | 0 | 0 | 0 | 0 | 0 | 21 | 23 | 4/7 | 3/3 |
| Author12-Article3 | 6% | 2 | 0 | 0 | 2 | 0 | 0 | 16 | 20 | 2/11 | 3/3 |
| Author12-Article5 | 5% | 0 | 0 | 4 | 0 | 0 | 0 | 19 | 23 | 3/5 | 3/3 |
| Author13-Article1 | 4% | 0 | 0 | 0 | 3 | 0 | 2 | 11 | 16 | 2/4 | 1/1 |
| Author14-Article1 | 2% | 0 | 0 | 0 | 0 | 0 | 0 | 9 | 9 | 3/7 | 1/2 |
| Author15-Article1 | 14% | 5 | 0 | 0 | 4 | 6 | 0 | 10 | 25 | 1/4 | 2/2 |
| Author15-Article2 | 9% | 0 | 0 | 0 | 0 | 0 | 0 | 0 | 0 | 1/5 | 2/2 |
| Author15-Article3 | 9% | 2 | 6 | 0 | 2 | 0 | 3 | 14 | 27 | 2/7 | 2/2 |
| Author15-Article5 | 19% | 0 | 6 | 0 | 29 | 7 | 3 | 23 | 68 | 2/4 | 2/2 |
| Author16-Article1 | 7% | 0 | 0 | 0 | 10 | 0 | 0 | 15 | 25 | 1/6 | 1/1 |
| Author17-Article2 | 3% | 0 | 0 | 0 | 0 | 0 | 0 | 6 | 6 | 1/3 | 1/1 |
| Author18-Article5 | 16% | 0 | 0 | 11 | 6 | 0 | 0 | 29 | 46 | 1/4 | 1/1 |
| Author19-Article1 | 0% | 0 | 0 | 0 | 0 | 0 | 0 | 0 | 0 | 2/3 | 1/1 |
| Author20-Article2 | 0% | 0 | 0 | 0 | 0 | 0 | 0 | 0 | 0 | 2/6 | 1/2 |
| Author21-Article2 | 5% | 5 | 0 | 0 | 0 | 5 | 1 | 23 | 34 | 4/8 | 1/2 |
| Author22-Article1 | 5% | 1 | 0 | 12 | 0 | 0 | 0 | 10 | 23 | 8/23 | 1/1 |
| Author22-Article2 | 8% | 0 | 0 | 5 | 0 | 0 | 0 | 44 | 49 | 2/8 | 1/1 |
| Author22-Article4 | 0% | 0 | 0 | 0 | 0 | 0 | 0 | 0 | 0 | 7/8 | 1/1 |
| Author22-Article5 | 0% | 0 | 0 | 0 | 0 | 0 | 0 | 0 | 0 | 2/8 | 1/1 |
| Author22-Article7 | 0% | 0 | 0 | 0 | 0 | 0 | 0 | 0 | 0 | 4/8 | 1/1 |
| Author22-Article9 | 2% | 0 | 0 | 0 | 0 | 0 | 0 | 2 | 2 | 2/4 | 1/1 |
| Author22-Article10 | 2% | 0 | 0 | 0 | 0 | 0 | 0 | 6 | 6 | 3/7 | 1/1 |
| Author22-Article12 | 11% | 0 | 0 | 0 | 26 | 0 | 0 | 13 | 39 | 2/6 | 1/1 |
| Author22-Article13 | 11% | 0 | 0 | 0 | 24 | 0 | 0 | 10 | 34 | 2/4 | 1/1 |
| Author22-Article16 | 3% | 1 | 0 | 0 | 0 | 0 | 0 | 11 | 12 | 5/7 | 1/1 |
| Author22-Article17 | 4% | 0 | 0 | 0 | 0 | 0 | 0 | 10 | 10 | 3/8 | 1/1 |

(Table 2: Non-derivative articles with supervisors as coauthors)

| AUTHORS-ARTICLES | SIMILARITY INDEX | TITLE | ABSTRACT | INTRODUCTION | METHODOLOGY | RESULTS | DISCUSSION | REFERENCES | MATCHES | AUTHOR POSITION | SUPERVISORS |
|---|---|---|---|---|---|---|---|---|---|---|---|
| Author22-Article19 | 2% | 0 | 0 | 0 | 0 | 0 | 0 | 7 | 7 | 2/8 | 1/1 |
| Author 23-Article1 | 5% | 1 | 0 | 0 | 0 | 0 | 0 | 14 | 15 | 3/5 | 1/2 |
| Author24-Article1 | 8% | 0 | 0 | 0 | 0 | 0 | 0 | 10 | 10 | 1/3 | 2/2 |
| Author24-Article2 | 6% | 0 | 0 | 0 | 0 | 0 | 0 | 22 | 22 | 3/8 | 2/2 |
| Author24-Article3 | 6% | 1 | 0 | 0 | 5 | 0 | 0 | 22 | 28 | 7/8 | 1/2 |
| Author24-Article5 | 0% | 0 | 0 | 0 | 0 | 0 | 0 | 0 | 0 | 3/6 | 1/2 |
| Author24-Article10 | 9% | 0 | 0 | 2 | 7 | 0 | 0 | 27 | 36 | 7/8 | 2/2 |
| Author24-Article11 | 9% | 0 | 0 | 3 | 14 | 0 | 0 | 12 | 29 | 1/4 | 2/2 |
| Author24-Article12 | 8% | 0 | 0 | 0 | 12 | 0 | 0 | 13 | 25 | 4/8 | 2/2 |
| Author25 -Article1 | 7% | 0 | 0 | 0 | 0 | 0 | 0 | 26 | 26 | 6/8 | 1/1 |
| Author26-Article1 | 4% | 0 | 0 | 0 | 0 | 0 | 0 | 18 | 18 | 4/8 | 1/2 |
| Author26-Article3 | 0% | 0 | 0 | 0 | 0 | 0 | 0 | 0 | 0 | 3/8 | 1/2 |
| Author28-Article1 | 1% | 0 | 0 | 0 | 0 | 0 | 0 | 2 | 2 | 2/4 | 1/2 |
| Author28-Article2 | 5% | 0 | 0 | 0 | 0 | 0 | 0 | 19 | 19 | 2/4 | 1/2 |
| Author28-Article5 | 12% | 0 | 0 | 0 | 14 | 0 | 0 | 22 | 36 | 3/4 | 1/2 |
| Author28-Article6 | 14% | 1 | 0 | 0 | 0 | 0 | 0 | 16 | 17 | 1/4 | 1/2 |
| Author29-Article1 | 22% | 1 | 0 | 3 | 6 | 0 | 0 | 0 | 10 | 1/4 | 2/2 |
| Author29-Article4 | 18% | 0 | 0 | 7 | 20 | 8 | 0 | 17 | 52 | 6/8 | 2/2 |
| Author29-Article5 | 0% | 0 | 0 | 0 | 0 | 0 | 0 | 0 | 0 | 1/4 | 1/2 |
| Author29-Article7 | 15% | 0 | 18 | 0 | 0 | 0 | 2 | 0 | 20 | 2/4 | 2/2 |
| Author29-Article8 | 5% | 0 | 0 | 1 | 4 | 2 | 0 | 7 | 14 | 4/8 | 2/2 |
| Author31-Article1 | 9% | 1 | 0 | 0 | 7 | 3 | 0 | 9 | 20 | 2/15 | 2/2 |
| Author32-Article1 | 7% | 0 | 0 | 0 | 32 | 14 | 0 | 62 | 108 | 2/5 | 1/1 |
| Author33-Article1 | 0% | 0 | 0 | 0 | 0 | 0 | 0 | 0 | 0 | 1/3 | 1/2 |
| Author33-Article2 | 0% | 0 | 0 | 0 | 0 | 0 | 0 | 0 | 0 | 3/5 | 1/2 |
| Author33-Article3 | 2% | 0 | 0 | 0 | 0 | 0 | 0 | 3 | 3 | 1/2 | 1/2 |
| Author33-Article4 | 0% | 0 | 0 | 0 | 0 | 0 | 0 | 0 | 0 | 2/5 | 1/2 |
| Author33-Article5 | 2% | 1 | 0 | 0 | 0 | 0 | 0 | 3 | 4 | 3/8 | 1/2 |
| Author34-Article1 | 9% | 1 | 0 | 15 | 0 | 0 | 0 | 12 | 28 | 1/8 | 1/2 |
| Author34-Article3 | 2% | 1 | 1 | 0 | 0 | 0 | 0 | 3 | 5 | 1/6 | 2/2 |
| Author35-Article1 | 30% | 1 | 0 | 4 | 32 | 46 | 6 | 31 | 120 | 1/4 | 2/2 |
| Author36-Article1 | 9% | 0 | 0 | 3 | 8 | 15 | 1 | 21 | 48 | 1/15 | 2/2 |
| Author36-Article3 | 6% | 0 | 0 | 0 | 0 | 0 | 0 | 21 | 21 | 3/8 | 2/2 |
| Author37-Article1 | 0% | 0 | 0 | 0 | 0 | 0 | 0 | 0 | 0 | 2/4 | 1/2 |
| Author37-Article2 | 0% | 0 | 0 | 0 | 0 | 0 | 0 | 0 | 0 | 1/4 | 1/2 |
| Author37-Article3 | 0% | 0 | 0 | 0 | 0 | 0 | 0 | 0 | 0 | 2/4 | 1/2 |
| Author37-Article6 | 11% | 0 | 3 | 1 | 0 | 0 | 2 | 38 | 44 | 1/5 | 2/2 |
| Author37-Article7 | 2% | 0 | 0 | 2 | 4 | 0 | 0 | 4 | 10 | 3/4 | 2/2 |
| Author37-Article8 | 0% | 0 | 0 | 0 | 0 | 0 | 0 | 0 | 0 | 1/7 | 1/2 |
| Author37-Article10 | 2% | 0 | 0 | 0 | 16 | 0 | 0 | 1 | 17 | 1/3 | 1/2 |
| Author38-Article1 | 11% | 0 | 0 | 0 | 0 | 0 | 3 | 49 | 52 | 4/6 | 1/2 |
| Author39-Article1 | 4% | 0 | 0 | 3 | 0 | 0 | 3 | 14 | 20 | 4/6 | 1/2 |
| Author40-Article1 | 16% | 0 | 0 | 3 | 9 | 35 | 0 | 13 | 60 | 1/4 | 1/1 |
| Author40-Article2 | 0% | 0 | 0 | 0 | 0 | 0 | 0 | 0 | 0 | 6/8 | 1/1 |
| Author41-Article1 | 7% | 0 | 2 | 0 | 11 | 0 | 0 | 10 | 23 | 2/5 | 1/2 |
| Author42-Article1 | 89% | 1 | 0 | 2 | 6 | 30 | 1 | 10 | 50 | 1/8 | 1/2 |
| Author42-Article2 | 4% | 1 | 0 | 5 | 3 | 0 | 0 | 11 | 20 | 3/8 | 1/2 |
| Author43-Article1 | 66% | 1 | 5 | 13 | 22 | 53 | 3 | 47 | 144 | 1/4 | 1/1 |
| Author44-Article4 | 10% | 1 | 0 | 6 | 0 | 0 | 0 | 9 | 16 | 5/6 | 2/2 |
| Author46-Article1 | 1% | 4 | 0 | 0 | 0 | 0 | 0 | 17 | 21 | 3/4 | 1/1 |
| Author46-Article2 | 8% | 0 | 0 | 2 | 0 | 0 | 0 | 9 | 11 | 1/4 | 1/1 |
| Author46-Article4 | 3% | 1 | 0 | 0 | 10 | 5 | 2 | 7 | 25 | 2/4 | 1/1 |
| Author46-Article7 | 5% | 0 | 0 | 0 | 0 | 0 | 0 | 7 | 7 | 5/6 | 1/1 |
| Author48-Article1 | 15% | 2 | 0 | 7 | 14 | 14 | 2 | 20 | 59 | 2/19 | 3/3 |
| Author50-Article1 | 2% | 0 | 0 | 0 | 3 | 7 | 2 | 7 | 19 | 1/5 | 2/2 |
| Author51-Article2 | 6% | 0 | 0 | 1 | 6 | 6 | 0 | 23 | 36 | 5/7 | 1/1 |
| Author51-Article4 | 9% | 0 | 2 | 0 | 0 | 1 | 0 | 32 | 35 | 1/5 | 1/1 |

(Table 2: Non-derivative articles with supervisors as coauthors)

| AUTHORS-ARTICLES | SIMILARITY INDEX | TITLE | ABSTRACT | INTRODUCTION | METHODOLOGY | RESULTS | DISCUSSION | REFERENCES | MATCHES | AUTHOR POSITION | SUPERVISORS |
|---|---|---|---|---|---|---|---|---|---|---|---|
| Author2-Article2 | 0% | 0 | 0 | 0 | 0 | 0 | 0 | 0 | 0 | 6/8 | 0/1 |
| Author6-Article1 | 0% | 0 | 0 | 0 | 0 | 0 | 0 | 0 | 0 | 11/29 | 0/2 |
| Author6-Article2 | 0% | 0 | 0 | 0 | 0 | 0 | 0 | 0 | 0 | 2/6 | 0/2 |
| Author6-Article3 | 0% | 0 | 0 | 0 | 0 | 0 | 0 | 0 | 0 | 9/16 | 0/2 |
| Author6-Article4 | 0% | 0 | 0 | 0 | 0 | 0 | 0 | 0 | 0 | 5/8 | 0/2 |
| Author7-Article4 | 0% | 0 | 0 | 0 | 0 | 0 | 0 | 0 | 0 | 4/4 | 0/2 |
| Author7-Article8 | 16% | 0 | 0 | 8 | 20 | 7 | 4 | 23 | 62 | 1/4 | 0/2 |
| Author9-Article5 | 0% | 0 | 0 | 0 | 0 | 0 | 0 | 0 | 0 | 3/5 | 0/1 |
| Author9-Article6 | 0% | 0 | 0 | 0 | 0 | 0 | 0 | 0 | 0 | 3/4 | 0/1 |
| Author12-Article4 | 3% | 0 | 0 | 0 | 0 | 0 | 0 | 10 | 10 | 1/4 | 0/3 |
| Author12-Article6 | 1% | 0 | 0 | 0 | 1 | 0 | 0 | 0 | 1 | 1/4 | 0/3 |
| Author15-Article4 | 0% | 0 | 0 | 0 | 0 | 0 | 0 | 0 | 0 | 1/4 | 0/2 |
| Author15-Article7 | 0% | 0 | 0 | 0 | 0 | 0 | 0 | 0 | 0 | 4/7 | 0/2 |
| Author17-Article1 | 0% | 0 | 0 | 0 | 0 | 0 | 0 | 0 | 0 | 4/8 | 0/1 |
| Author18-Article1 | 0% | 0 | 0 | 0 | 1 | 0 | 0 | 0 | 1 | 4/6 | 0/1 |
| Author18-Article3 | 0% | 0 | 0 | 0 | 0 | 0 | 0 | 0 | 0 | 4/7 | 0/1 |
| Author18-Article4 | 0% | 0 | 0 | 0 | 0 | 0 | 0 | 0 | 0 | 1/3 | 0/1 |
| Author18-Article6 | 0% | 0 | 0 | 0 | 0 | 0 | 0 | 0 | 0 | 6/13 | 0/1 |
| Author18-Article7 | 0% | 0 | 0 | 0 | 0 | 0 | 0 | 0 | 0 | 2/4 | 0/1 |
| Author20-Article1 | 0% | 0 | 0 | 0 | 0 | 0 | 0 | 0 | 0 | 2/5 | 0/2 |
| Author22-Article3 | 2% | 0 | 0 | 0 | 0 | 0 | 0 | 8 | 8 | 3/8 | 0/1 |
| Author22-Article6 | 0% | 0 | 0 | 0 | 0 | 0 | 0 | 0 | 0 | 4/11 | 0/1 |
| Author22-Article8 | 0% | 0 | 0 | 0 | 0 | 0 | 0 | 0 | 0 | 4/7 | 0/1 |
| Author22-Article18 | 0% | 0 | 0 | 0 | 0 | 0 | 0 | 0 | 0 | 2/6 | 0/1 |
| Author24-Article8 | 4% | 0 | 0 | 0 | 0 | 0 | 0 | 17 | 17 | 2/5 | 0/2 |
| Author24-Article9 | 5% | 0 | 0 | 0 | 5 | 0 | 0 | 16 | 21 | 2/5 | 0/2 |
| Author26-Article2 | 0% | 0 | 0 | 0 | 0 | 0 | 0 | 0 | 0 | 3/6 | 0/2 |
| Author29-Article2 | 0% | 0 | 0 | 0 | 0 | 0 | 0 | 0 | 0 | 2/2 | 0/2 |
| Author29-Article6 | 0% | 0 | 0 | 0 | 0 | 0 | 0 | 0 | 0 | 1/3 | 0/2 |
| Author30-Article1 | 4% | 1 | 0 | 0 | 0 | 0 | 4 | 15 | 20 | 6/7 | 0/1 |
| Author30-Article2 | 0% | 0 | 0 | 0 | 0 | 0 | 0 | 0 | 0 | 1/3 | 0/1 |
| Author37-Article4 | 0% | 0 | 0 | 0 | 0 | 0 | 0 | 0 | 0 | 2/3 | 0/2 |
| Author37-Article9 | 0% | 0 | 0 | 0 | 0 | 0 | 0 | 0 | 0 | 1/3 | 0/2 |
| Author39-Article2 | 4% | 1 | 0 | 3 | 1 | 0 | 0 | 10 | 15 | 2/6 | 0/2 |
| Author44-Article2 | 1% | 0 | 0 | 2 | 0 | 0 | 0 | 7 | 9 | 2/6 | 0/2 |
| Author45-Article1 | 0% | 0 | 0 | 0 | 0 | 0 | 0 | 0 | 0 | 1/6 | 0/2 |
| Author45-Article2 | 0% | 0 | 0 | 0 | 0 | 0 | 0 | 0 | 0 | 1/5 | 0/2 |
| Author45-Article3 | 0% | 0 | 0 | 0 | 0 | 0 | 0 | 0 | 0 | 6/9 | 0/2 |
| Author45-Article4 | 0% | 0 | 0 | 0 | 0 | 0 | 0 | 0 | 0 | 5/6 | 0/2 |
| Author45-Article5 | 0% | 0 | 0 | 0 | 0 | 0 | 0 | 0 | 0 | 5/6 | 0/2 |
| Author45-Article6 | 1% | 0 | 0 | 0 | 0 | 0 | 0 | 3 | 3 | 6/11 | 0/2 |
| Author45-Article7 | 0% | 0 | 0 | 0 | 0 | 0 | 0 | 0 | 0 | 2/6 | 0/2 |
| Author45-Article8 | 0% | 0 | 0 | 0 | 0 | 0 | 0 | 0 | 0 | 1/6 | 0/2 |
| Author45-Article9 | 0% | 0 | 0 | 0 | 0 | 0 | 0 | 0 | 0 | 1/7 | 0/2 |
| Author45-Article10 | 0% | 0 | 0 | 0 | 0 | 0 | 0 | 0 | 0 | 1/9 | 0/2 |
| Author46-Article6 | 6% | 0 | 0 | 0 | 3 | 11 | 3 | 12 | 29 | 3/5 | 0/1 |
| Author47-Article1 | 0% | 0 | 0 | 0 | 0 | 0 | 0 | 0 | 0 | 1/5 | 0/3 |
| Author47-Article2 | 0% | 0 | 0 | 0 | 0 | 0 | 0 | 0 | 0 | 2/5 | 0/3 |
| Author47-Article3 | 0% | 0 | 0 | 0 | 0 | 0 | 0 | 0 | 0 | 1/9 | 0/3 |
| Author47-Article4 | 0% | 0 | 0 | 0 | 0 | 0 | 0 | 0 | 0 | 7/14 | 0/3 |
| Author47-Article5 | 3% | 0 | 0 | 0 | 0 | 0 | 0 | 6 | 6 | 5/7 | 0/3 |
| Author47-Article6 | 0% | 0 | 0 | 0 | 0 | 0 | 0 | 0 | 0 | 3/4 | 0/3 |
| Author49-Article1 | 0% | 0 | 0 | 0 | 0 | 0 | 0 | 0 | 0 | 2/6 | 0/1 |

(Table 3: Non-derivative articles without theses supervisors as coauthors)